\documentclass[11pt,a4paper]{article}
\usepackage{amsfonts,amsmath,amsthm}
\usepackage{amstext}
\usepackage[pdftex]{graphicx, color}
\usepackage{longtable}

\graphicspath{{figures/}}
\graphicspath{{re/}}

  {\par\medbreak\refstepcounter{theorem}%
    \noindent\textbf{Remark~\thetheorem. }}%
  {\par\medskip}

\newcommand{\vz}[1]{\ensuremath{\mathbb{#1}}}

\newcommand{\R}{{\vz R}}

\long\def\drop#1{}

\let\epsilon\varepsilon

\def\XXint#1#2#3{{\setbox0=\hbox{$#1{#2#3}{\int}$}
     \vcenter{\hbox{$#2#3$}}\kern-.5\wd0}}

\usepackage{geometry}                
\usepackage{graphicx}
\usepackage{amssymb}
\usepackage{epstopdf}
\usepackage{latexsym}
\usepackage{amssymb}
\usepackage{amsfonts}
\usepackage{bm}
\usepackage{epsfig}
\usepackage{bbm}
\usepackage{amsmath}

\usepackage{psfrag}
\usepackage{color}
\usepackage{multicol}
\usepackage{subfig}
\usepackage{appendix}
\usepackage{relsize}
\usepackage{rotating}
\usepackage{color, colortbl}
\usepackage{morefloats}

\definecolor{Gray}{gray}{0.9}

\newcommand{\footnoteremember}[2]{
\footnote{#2}
\newcounter{#1}
\setcounter{#1}{\value{footnote}}
}
\newcommand{\footnoterecall}[1]{
\footnotemark[\value{#1}]
}

\title{Community detection using spectral clustering on sparse geosocial data}

\date{}                                           

\author{Yves van Gennip\footnoteremember{UCLAmath}{Department of Mathematics, Applied Mathematics, UCLA}, Blake Hunter\footnoterecall{UCLAmath},\\ Raymond Ahn \footnote{Department of Mathematics, CSULB}, Peter Elliott\footnoterecall{UCLAmath}, Kyle Luh\footnote{Department of Physics, Yale},\\Megan Halvorson\footnoteremember{UCISoc}{School of Social Ecology, Department of Criminology, Law and Society, UCI}, Shannon Reid\footnoterecall{UCISoc}, Matthew Valasik\footnoterecall{UCISoc}, James Wo\footnoterecall{UCISoc},\\George E. Tita\footnoterecall{UCISoc}, Andrea L. Bertozzi\footnoterecall{UCLAmath}, P. Jeffrey Brantingham\footnote{Department of Anthropology, UCLA}}

\begin{document}

\maketitle

\begin{abstract}
In this article we identify social communities among gang members in the Hollenbeck policing district in Los Angeles, based on sparse observations of a combination of social interactions and geographic locations of the individuals. This information, coming from LAPD Field Interview cards, is used to construct a similarity graph for the individuals. We use spectral clustering to identify clusters in the graph, corresponding to communities in Hollenbeck, and compare these with the LAPD's knowledge of the individuals' gang membership. We discuss different ways of encoding the geosocial information using a graph structure and the influence on the resulting clusterings. Finally we analyze the robustness of this technique with respect to noisy and incomplete data, thereby providing suggestions about the relative importance of quantity versus quality of collected data. 
\end{abstract}
\textbf{Keywords:} spectral clustering, stability analysis, social networks, community detection, data clustering, street gangs, rank-one matrix update\\
\textbf{MSC 2010:} 62H30, 91C20, 91D30, 94C15

\bigskip

\section{Introduction}
Determining the communities into which people organize themselves is an important step towards understanding their behavior. In diverse contexts, from advertising to risk assessment, the social group to which someone belongs can reveal crucial information. 
In practical situations only limited information is available to determine these communities. Peoples' geographic location at a set of sample times is often known, but it may be asked whether this provides enough information for reliable 
community detection. 
In many situations social interactions also can be inferred, from observing people in the same place at the same time. This information can be very sparse.
The question is how to get the most community information out of these limited observations.
Here we show that social communities within a group of street gang members can be detected by complementing sparse (in time) geographical information with imperfect, but not too sparse, knowledge of the social interactions. 
First we construct a graph from LAPD Field Interview (FI) card information about individuals in the Hollenbeck policing area of Los Angeles, which has a high density of street gangs. The nodes represent individuals and the edges between them are weighted according to their geosocial similarity. When using this extremely sparse social data in combination with the geographical data, the eigenvectors of the graph display hotspots at major gang locations. However, the available collected social data is too sparse and the social situation in Hollenbeck too complex (communities do not necessarily proxy for gang boundaries) for the resulting clustering, constructed using the spectral clustering algorithm, to identify gangs accurately. Extending the available social data past the current sparsity level by artificially adding (noisy) ground truth consisting of true connections between members of the same gang leads to quantitative improvements of clustering metrics.
This shows that limited information about peoples' whereabouts and interactions can suffice to determine which social groups they belong to, but the allowed sparsity in the social data has its limits. However, no detailed personal information or knowledge about the contents of their interactions is needed. The sparsity in time of the geographical information is mitigated by the relative stability in time of the gang territories. 

The case of criminal street gangs speaks to a more general social group classification problem found in both security- and non-security-related contexts.  In an active insurgency, for example, the human terrain contains individuals from numerous family, tribal and religious groups.  The border regions of Afghanistan are home to perhaps two dozen distinct ethno-linguistic groups and many more family and tribal organizations \cite{JohnsonMason08}.  Only a small fraction of the individuals are actively belligerent, but many may passively support the insurgency.  Since support for an insurgency is related in part to family, tribal and religious group affiliations, as well as more general social and economic grievances \cite{Kilcullen09}, being able to correctly classify individuals to their affiliated social groups may be extremely valuable for isolating and impacting hostile actors.  Yet, on-the-ground intelligence is difficult to collect in extreme security settings.  While detailed individual-level intelligence may not be readily available, observations of where and with whom groups of individuals meet may indeed be possible.  The methods developed here may find application in such contexts.

In non-security contexts, establishing an individualÕs group affiliation and, more broadly, the structure of a social group can be extremely costly, requiring detailed survey data collection.  Since much routine social and economic activity is driven by group affiliation \cite{ChenLi09}, lower cost alternatives to group classification may be valuable for encouraging certain types of behavior.  For example, geotagged social media activity, such as Facebook, Twitter or Instagram posts, might reveal the geo-social context of individual activities \cite{WatanabeOchiOkabeOnai11}.  The methods developed here could be used to establish group affiliations of individuals under these circumstances.

This paper applies spectral clustering to an interesting new street gang data set. We study how social and geographical data can be combined to have the resulting clusters approximate existing communities in Hollenbeck, and investigate the limitations of the method due to the sparsity in the social data.

\section{The setting}

\begin{figure}[ht]
	\begin{center}
		$\begin{array}{c@{\hspace{.1in}}c}
	\includegraphics[width=.5\textwidth]{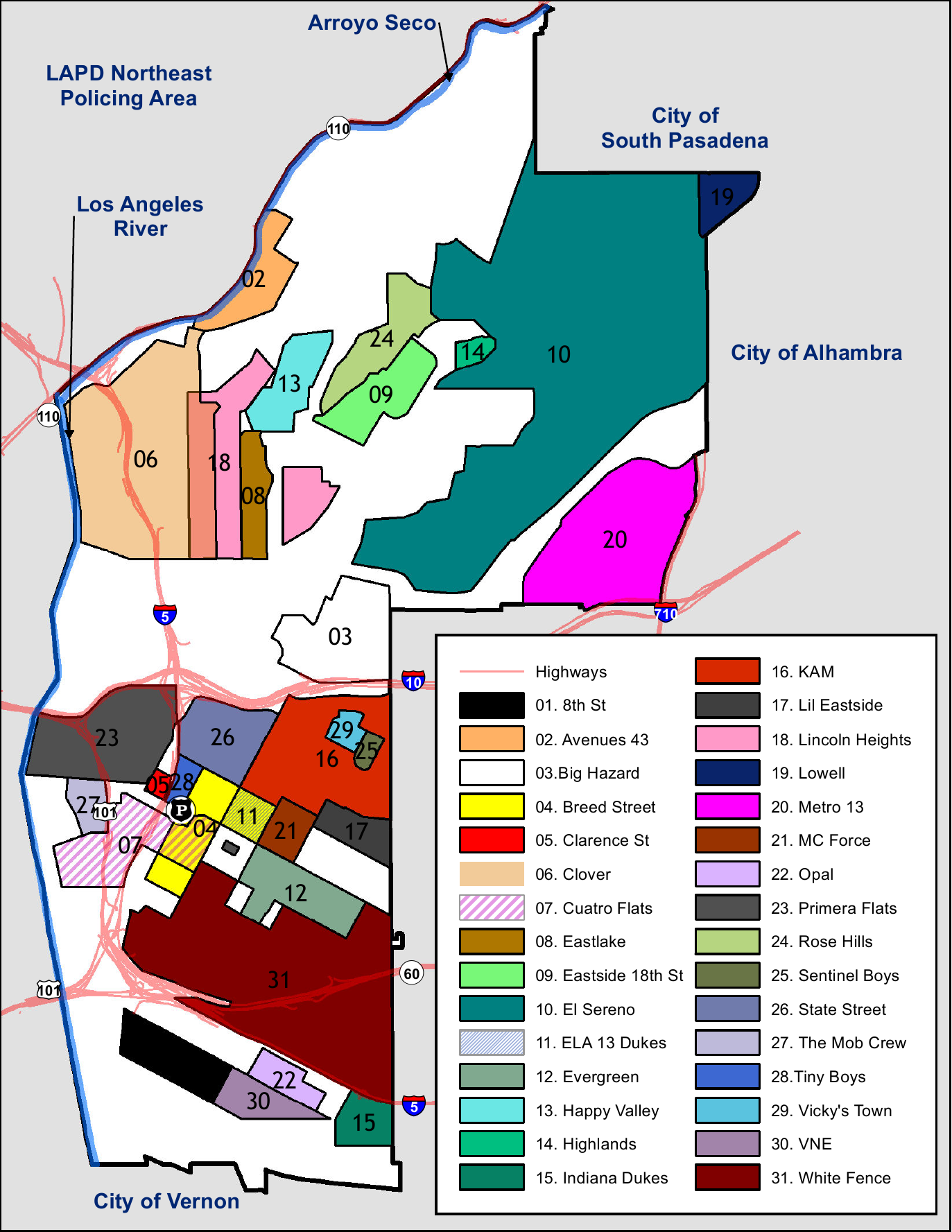} &  \includegraphics[width=.509\textwidth]{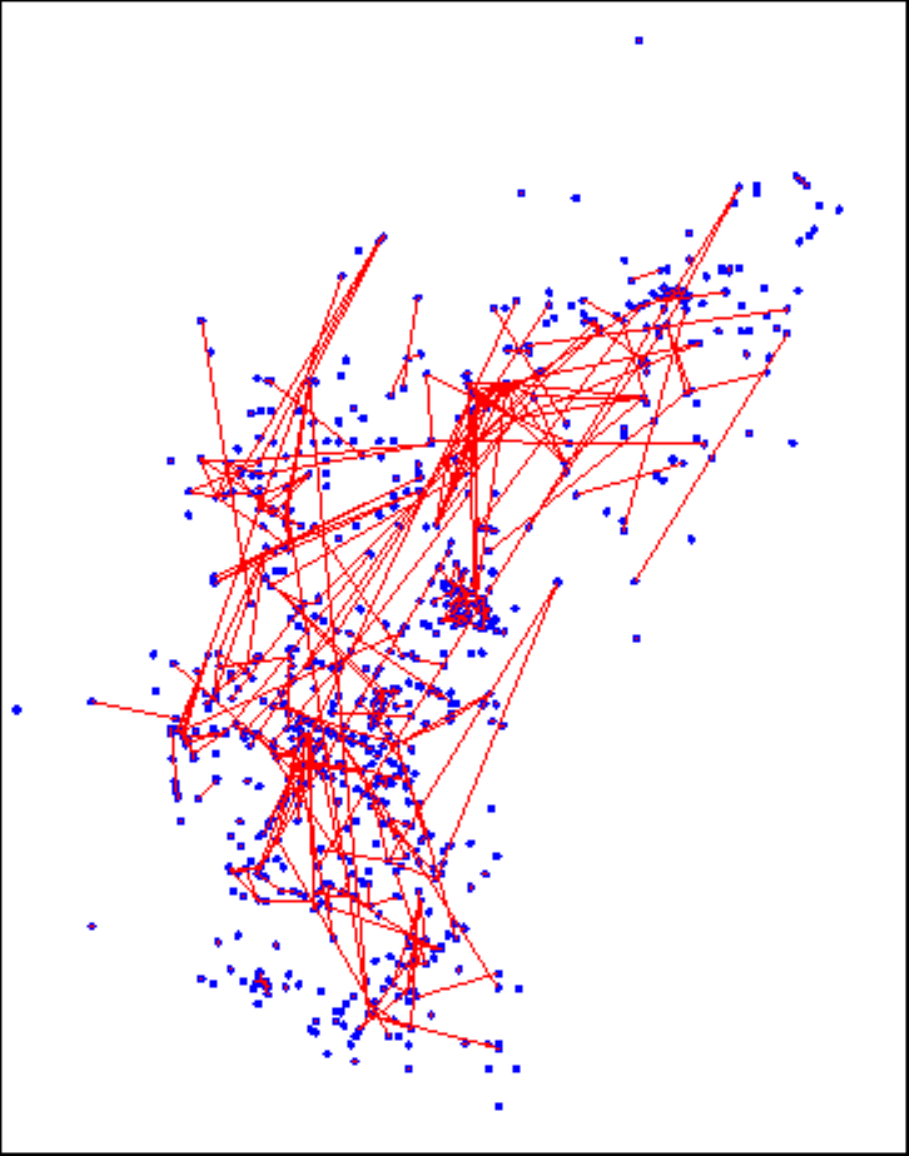}
		\end{array}$
	
	\end{center}
	\caption {\footnotesize  \textbf{Left :} Map of gang territories in the Hollenbeck area of Los Angeles.  \textbf{Right:}  LAPD FI card data showing average stop location of 748 individuals with social links of who was stopped with whom.}\label{fig:Hollenbeckdata}
\end{figure}

Hollenbeck (Figure~\ref{fig:Hollenbeckdata}, left) is bordered by the Los Angeles River, the Pasadena Freeway and areas which do not have rivaling street gangs \cite{RadilFlintTita10}. The built and and natural boundaries sequester Hollenbeck's gangs from neighboring communities, inhibiting socialization. In recent years quite a few sociological, \textit{e.g.} \cite{TitaRileyRidgewayGrammichAbrahamseGreenwood03,RadilFlintTita10,TitaRadil11} and mathematical papers, \textit{e.g.} \cite{HegemannSmithBarbaroBertozziReidTita11,MohlerShortBrantinghamSchoenbergTita11,HegemannLewisBertozzi12,ShortMohlerBrantinghamTita12}, on the Hollenbeck gangs have been produced, but none in the area of gang clustering. 

The recent social science/policy research on Hollenbeck gangs has combined both the geographic and social position of gangs to better understand the relational nature of gang violence.  Clustering gangs both in terms of their spatial adjacency and position in a rivalry network has shown that structurally equivalent \cite{WassermanFaust94} gangs experience similar levels of violence \cite{RadilFlintTita10}.  Incorporating both the social and geographical distance into contagion models of gang violence provides a more robust analysis \cite{TitaRadil11}.  Additionally, ecological models of foraging behavior have shown that even low levels of inter-gang competition produce sharply delineated boundaries among gangs with violence following predictable patterns along these borders \cite{BrantinghamTitaShortReid12}.  Accounting for these socio-spatial dimensions of gang rivalries has contributed to the design of successful interventions aimed at reducing gun violence committed by gangs \cite{TitaRileyRidgewayGrammichAbrahamseGreenwood03}.  An evaluation of this intervention demonstrated that geographically targeted enforcement of two gangs reduced gun violence in the focal neighborhoods. The crime reduction benefits also diffused through the social network as the levels of violence among the targeted gangsÕ rivals also decreased.

In this article we use one year's worth (2009) of LAPD FI cards. These cards are created at the officer's discretion whenever an interaction occurs with a civilian. They are not restricted to criminal events. Our data set is restricted to FI cards concerning stops involving known or suspected Hollenbeck gang members\footnote{In the FI card data set for some individuals certain data entries were missing. We did not include these individuals in our data set either.}. We further restricted our data set to include only the 748 individuals (anonymized) whose gang affiliation is recorded in the FI card data set (based on expert knowledge). These affiliations serve as a ground truth for clustering. From each individual we use information about the average of the locations where they were stopped and which other individuals were present at each stop (Figure~\ref{fig:Hollenbeckdata}, right) in our algorithm.

\section{The method}

We construct a fully connected graph whose nodes represent the 748 individuals. Every pair of nodes $i$ and $j$ is connected by an edge with weight
\[
W_{i,j}  = \alpha S_{i,j} + (1-\alpha) e^{-d_{i,j}^2/\sigma^2},
\]
where $\alpha\in [0,1]$, $d_{i,j}$ is the standard Euclidean distance between the average stop locations of individuals $i$ and $j$, and $\sigma$ is chosen to be the length which is one standard deviation larger than the mean distance between two individuals who have been stopped together\footnote{Most results in this paper are fairly robust to small perturbations that keep $\sigma$ of the same order of magnitude ($10^3$ feet), \textit{e.g.} replacing it by just the mean distance. The mean distance between members of the same gang (computed using the ground truth) is of the same order of magnitude. Another option one could consider, is to use local scaling, such that $\sigma$ has a different value for each pair $i, j$, as in \cite{Zelnik-ManorPerona04}. We will not pursue that approach here. Our focus will be mainly on the roles of $\alpha$ and $S_{i,j}$.}. The choice of Gaussian kernel for the geographic distance dependent part of $W$ is a natural one (since it models a diffusion process) setting the width of the kernel to be the length scale within which most social interactions take place. 
We encode social similarity by taking $S=A$, where $A$ is the social adjacency matrix with entry $A_{i,j}=1$ if $i$ and $j$ were stopped together (or $i=j$) and $A_{i,j}=0$ otherwise. In Section~\ref{sec:socialmatrix} we discuss some other choices for $S$ and how the results are influenced by their choice. Note that, because of the typically non-violent nature of the stops, we assume that individuals that were stopped together share a friendly social connection, thus establishing a social similarity link. The parameter $\alpha$ can be adjusted to set the relative importance between social and geographic information. If $\alpha = 0$ only geographical information is used, if $\alpha=1$ only social information.

Using spectral clustering (explained below) we group the individuals into $31$ different clusters. The modeling assumption is that these clusters correspond to social communities among Hollenbeck gang members. We study the question how much these clusters or communities resemble the actual gangs, as defined by each individual's gang affiliation given on the FI cards. The \textit{a priori} choice for 31 clusters is motivated by the LAPD's observation that there were 31 active gangs in Hollenbeck at the time the data was collected, each of which is represented in the data set\footnote{The number of members of each gang in the data set varies between 2 and 90, with an average of 24.13 and a standard deviation of 21.99.}. In Appendix~\ref{app:different} we briefly discuss some results obtained for different values of $k$. The question whether this number can be deduced from the data without prior assumption ---and if not, what that means for either the data or the LAPD's assumption--- is both mathematically and anthropologically relevant, but falls mostly outside the scope of this paper. It is partly addressed in current work \cite{HuvanGennipHunterBertozziPorter12,vanGennipHuHunterPorter12} that uses the modularity optimization method (possibly with resolution parameter) (\cite{NewmanGirvan04,Newman06,PorterOnnelaMucha09} and references therein), and its extension, the multislice modularity minimization method of \cite{MuchaRichardsonMaconPorterOnnela10}. We stress that our method clusters the individuals into 31 sharply defined clusters. Other methods are available to find mixed-membership communities \cite{KoutsourelakisEliassi-Rad08,Eliassi-RadHenderson10}, but we will not pursue those here.

We use a spectral clustering algorithm \cite{NgJordanWeiss02} for its simplicity and transparency in making non-separable (\text{i.e.} not linearly separable) clusters separable. At the end of this paper we will discuss some other methods that can be used in future studies.

We compute the matrix $V$, whose columns are the first $31$ eigenvectors (ordered according to decreasing eigenvalues) of the normalized affinity matrix $D^{-1} W$. Here $D$ is a diagonal matrix with the nodes' degrees on the diagonal: $D_{i,i} := \sum_{j=1}^{748} W_{i,j}$. These eigenvectors are known to solve a relaxation of the normalized cut (Ncut) problem \cite{ShiMalik00,YuShi03,VonLuxburg07}, by giving non-binary approximations to indicator functions for the clusters.  
We turn them into binary approximations using the $k$-means algorithm \cite{HartiganWong79} on the rows of $V$. Note that each row corresponds to an individual in the data set and assigns it a coordinate in $\R^{31}$. The $k$-means algorithm iteratively assigns individuals to their nearest centroid and updates the centroids after each step. Because $k$-means uses a random initial seeding of centroids, in the computation of the metrics below we average over 10 $k$-means runs. 

We investigate two main questions. The first is sociological: Is it possible to identify social structures in human behavior from limited observations of locations and colocations of individuals and how much does each data type contribute? Specifically, do we benefit from adding geographic data to the social data? We also look at how well our specific FI card data set performs in this regard. The second question is essentially a modeling question: How should we choose $\alpha$ and $S$ to get the most information out of our data, given that our goal is to identify gang membership of the individuals in our data set? Hence we compute metrics comparing our clustering results to the known gang affiliations and investigate the stability of these metrics for different modeling choices.

\section{The metrics}\label{sec:metrics}

We focus primarily on a purity metric and the $z$-Rand score, which are used to compare two given clusterings. For purity one of the clusterings has to be assigned as the true clustering, this is not necessary for the $z$-Rand score. In Appendix~\ref{app:othermetrics} we discuss other metrics and their results.

Purity is an often used clustering metric, \textit{e.g.} \cite{HarrisAubertHaeb-UmbachBeyerlein1999}. It is the percentage of correctly classified individuals, when classifying each cluster as the gang in the majority in that cluster (in the case of a tie any of the majority gangs can be chosen, without affecting the purity score). Note that we allow multiple clusters to be classified as the same gang.

To define the $z$-Rand score we first need to introduce the pair counting quantity\footnote{Not to be confused with the matrix element $W_{1,1}$.} $w_{11}$, which is the number of pairs which belong both to the same cluster in our $k$-means clustering (say, clustering $A$) and to the same gang according the ``ground truth'' FI card entry (say, clustering $B$), \textit{e.g.} \cite{Meila07,TraudKelsicMuchaPorter11} and references therein. The $z$-Rand score $z_R$, \cite{TraudKelsicMuchaPorter11}, is the number of standard deviations which $w_{11}$ is removed from its mean value under a hypergeometric distribution of equally likely assignments subject to new clusterings $\hat A$ and $\hat B$ having the same numbers and sizes of clusters as clusterings $A$ and $B$, respectively. 

Note that purity is a measure of the number of correctly classified {\it individuals}, while the $z$-Rand score measures correctly identified {\it pairs}. Purity thus has a bias in favor of more clusters. In the extreme case in which each individual is assigned to its own cluster (in clustering $A$), the purity score is 100\%. However, in this case the number of correctly identified pairs is zero (each gang in our data set has at least two members), and the mean and standard deviation of the hypergeometric distribution are zero. Hence the $z$-Rand score is not well-defined. At the opposite extreme, where we cluster all individuals into one cluster in clustering $A$, we have the maximum number of correctly classified pairs, but the standard deviation of the hypergeometric distribution is again zero, hence the $z$-Rand score is again not well-defined. The $z$-Rand score thus automatically shows warning signs in these extreme cases. Slight perturbations from these extremes will have very low $z$-Rand scores, and hence will also be rated poorly by this metric. Since we prescribe the number of clusters to be 31, this bias of the purity metric will not play an important role in this paper.

As a reference to compare the results discussed in the next section to, the total possible number of pairs among the 748 individuals is 279,378. Of these pairs, 15,904 involve members of the same gang, and 263,474 pairs involve members of different gangs (according to the ground truth). The $z$-Rand score for the clustering into true gangs is 404.7023.

\section{Performance of FI card data set}\label{sec:performance}

In Table~\ref{Table:S=A_1} we show the purity and $z$-Rand scores using $S=A$ for different $\alpha$ (for each $\alpha$ we give the average value over 10 $k$-means runs and the standard deviation). Clearly $\alpha=1$ is a bad choice. This is unsurprising given the sparsity of the social data. The clustering thus dramatically improves when we add geographical data to the social data.

On the other end of the spectrum $\alpha=0$ gives a purity that is within the error bars of the optimum value (at $\alpha=0.4$), indicating that a lot of the gang structure in Hollenbeck is determined by geography. This is not unexpected, given the territorial nature of these gangs. However, the $z$-Rand score can be significantly improved by choosing a nonzero $\alpha$ and hence again we see that a mix of social and geographical data is preferred.

\begin{table}[ht]
\begin{tabular}{|c|c|c|}
\hline
$\alpha$ & Purity & $z$-Rand\\
\hline
0 & 0.5548 $\pm$ 0.0078 & 120.6910 $\pm$ 19.4133
\\
0.1 & 0.5595 $\pm$ 0.0136 & 131.8397 $\pm$ 18.5551
\\
0.2 & 0.5574 $\pm$ 0.0100 & 121.9785 $\pm$ 18.3149
\\
0.3 & 0.5612 $\pm$ 0.0115 \cellcolor{Gray} & 137.2643 $\pm$ 21.0990
\\
0.4 & 0.5603 $\pm$ 0.0087 & 142.9746 $\pm$ 15.9186
\\
0.5 & 0.5531 $\pm$ 0.0118 & 139.8599 $\pm$ 14.2651
\\
0.6 & 0.5452 $\pm$ 0.0107 & 141.7835 $\pm$ 13.4852
\\
0.7 & 0.5452 $\pm$ 0.0099 & 130.2264 $\pm$ 21.5967
\\
0.8 & 0.5460 $\pm$ 0.0104 & 134.9519 $\pm$ 25.2803
\\
0.9 & 0.5602 $\pm$ 0.0061 & 145.7576 $\pm$ 13.4988 \cellcolor{Gray}
\\
1 & 0.2568 $\pm$ 0.0158 & 6.1518 $\pm$ 1.7494
\\
\hline
\end{tabular}
\caption{\footnotesize A list of the mean $\pm$ standard deviation over ten $k$-means runs of the purity and $z$-Rand score, using $S=A$. Cells with the optimal mean value are highlighted. Note however that other values are often close to the optimum compared to the standard deviation.}\label{Table:S=A_1}
\end{table}

In Appendix~\ref{app:othermetrics} we discuss the results we got from some other metrics, like ingroup homogeneity and outgroup heterogeneity measures and Hausdorff distance between the cluster centers. They show similar behavior as purity and the $z$-Rand score: All of them are limited by the sparsity and noisiness of the available data, but they typically show that it is preferable to include both social and geographical data. Especially social data by itself usually performs badly.

\begin{figure}[ht]
    \begin{center}
    $\begin{array}{c}
    \includegraphics[width=0.63\textwidth]{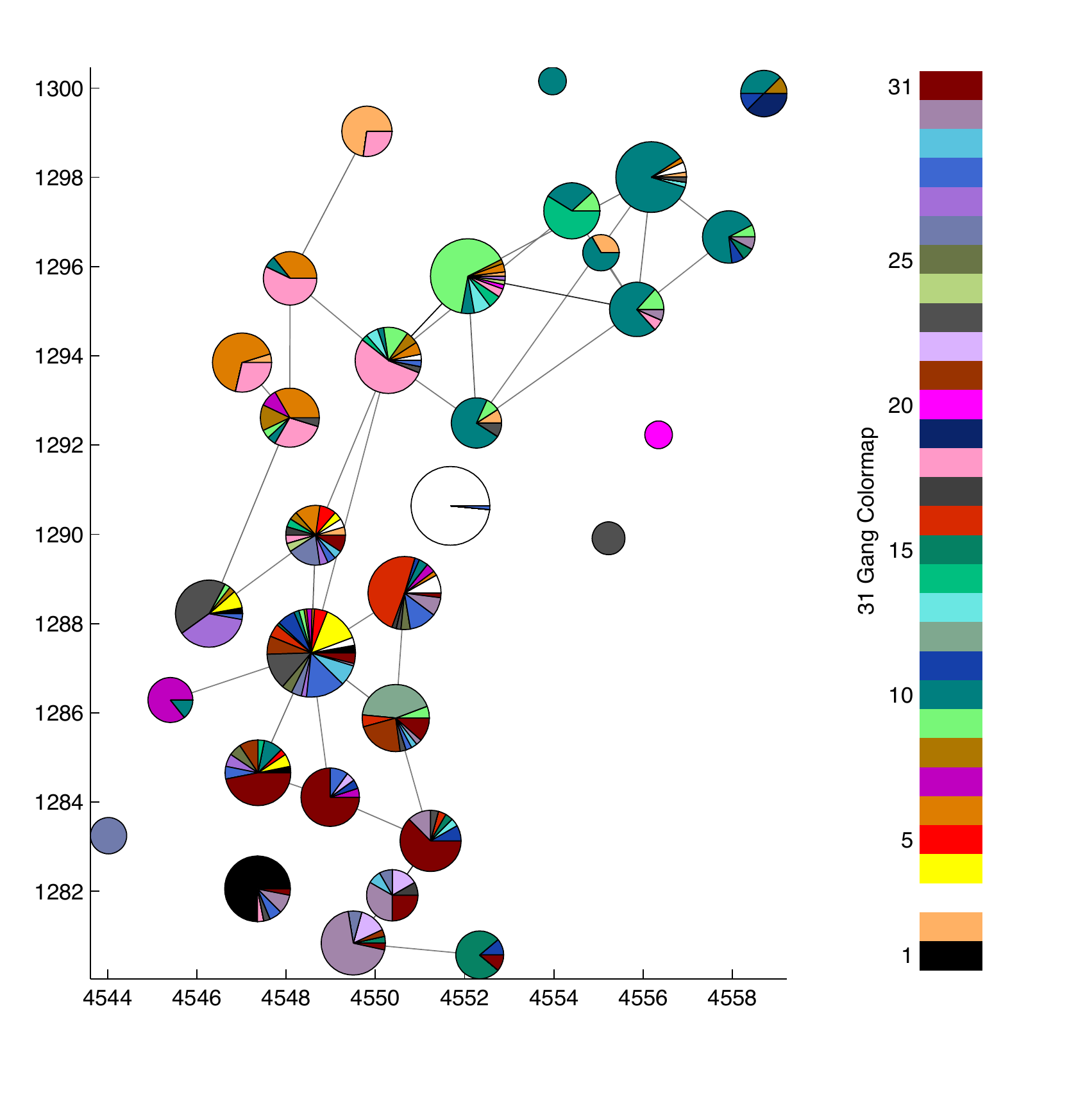}
    \end{array}$
    \end{center}
\caption{\footnotesize Pie charts made with code from \cite{TraudFrostMuchaPorter09} for a spectral clustering run with $S=A$ and $\alpha=0.4$. The size of each pie represents the cluster size and each pie is centered at the centroid of the average positions of the individuals in the cluster. The coloring indicates the gang make-up of the cluster and agrees with the gang colors in Figure~\ref{fig:Hollenbeckdata}. The legend shows the 31 different colors which are used, with the numbering of the gangs as in Figure~\ref{fig:Hollenbeckdata}. The axes are counted from an arbitrary but fixed origin. For aesthetic reasons the unit on both axes is approximately $435.42$ meters. The connections between pie charts indicate inter-cluster social connections (\textit{i.e.} nonzero elements of $A$).}\label{fig:piechartclustering}
\end{figure}

Figure~\ref{fig:piechartclustering} shows a pie chart (made with code from \cite{TraudFrostMuchaPorter09}) of one run of the spectral clustering algorithm, using $S=A$ and $\alpha=0.4$. We see that some clusters are quite homogeneous, especially the dark blue cluster located in Big Hazard's territory. Others are fragmented. We may interpret these results in light of previous work \cite{DeckerCurry02a}, which suggests that gangs vary substantially in their degree of internal organization. However, recall that in this paper we prescribe the number of clusters to be 31, so gang members are forced to cluster in ways that may not represent true gang organization.

Table~\ref{Table:S=A_1}, the pie charts in Figure~\ref{fig:piechartclustering}, and the other metrics discussed in Appendix~\ref{app:othermetrics} paint a consistent picture: The social data in the FI card data set is too sparse to stand on its own. Adding a little bit of geographic data however immensely improves the results. Geographic data by itself does pretty well, but can typically be improved by adding some social data. However, even for the optimal values the clustering is far from perfect. Therefore we will now consider different social matrices $S$ with two questions in mind: 1) Can we improve the performance of the social data by encoding it differently? 2) Is it really the sparsity of the social data that is the problem, or can the spectral clustering method not perform any better even if we would have more social data? The first question will be studied in Section~\ref{sec:socialmatrix}, the second in Section~\ref{sec:stability}.

\section{Different social matrices}\label{sec:socialmatrix}

For the results discussed above we have used the social adjacency matrix $A$ as the social matrix $S$. However, there are some interesting observations to make if we consider different choices for $S$.

The first alternative we consider is the social environment matrix $E$, which is a normalized measure of how many social contacts two individuals have in common. Its entries range between 0 and 1, a high value indicating that $i$ and $j$ met a lot of the same people (but, if $E_{i,j}<1$, not necessarily each other) and a low value indicating that $i$ and $j$'s social neighborhoods are (almost) disjoint. It is computed as follows. Let $f^i$ be the $i^{\text{th}}$ column of $A$. Then $E$ has entries $\displaystyle E_{i,j} = \sum_{k=1}^{748} \frac{f^i_k f^j_k}{\|f^i\| \|f^j\|}$ (where $\displaystyle \|f^i\|^2 = \sum_{k=1}^{748} (f^i_k)^2$). The procedure is reminiscent of the nonlocal means method \cite{BuadesCollMorel05} in image analysis, in which pixel patches are compared, instead of single pixels.

From our simulations (not listed here) we have seen that we get very similar results using either $S=A$ or $S=E$, both in terms of the optimal values for our metrics and whether these optima are achieved at the ends of the $\alpha$-interval (\textit{i.e.} $\alpha=0$ or $\alpha=1$) or in the interior ($0<\alpha<1$). The simulations described in Section~\ref{sec:stability} below showed that even for less sparse and more accurate data the results for $S=A$ and $S=E$ are similar.

\smallskip

An interesting visual phenomenon happens when, instead of using $A$ or $E$, we use a rank-one update of these matrices as the social matrix $S$. To be precise, we set $S= n (A+C)$ where $C$ is the matrix with $C_{i,j}=1$ for every entry and $n^{-1} := \underset{i,j}\max\, (A+C)_{i,j}$ is a normalization factor such that the maximum entry in $S$ is equal to $1$. (Again, the results are similar if we use $E$ instead of $A$.)

\begin{figure}[ht]
	\begin{center}
		$\begin{array}{c@{\hspace{.1in}}}
		\includegraphics[width=\textwidth]{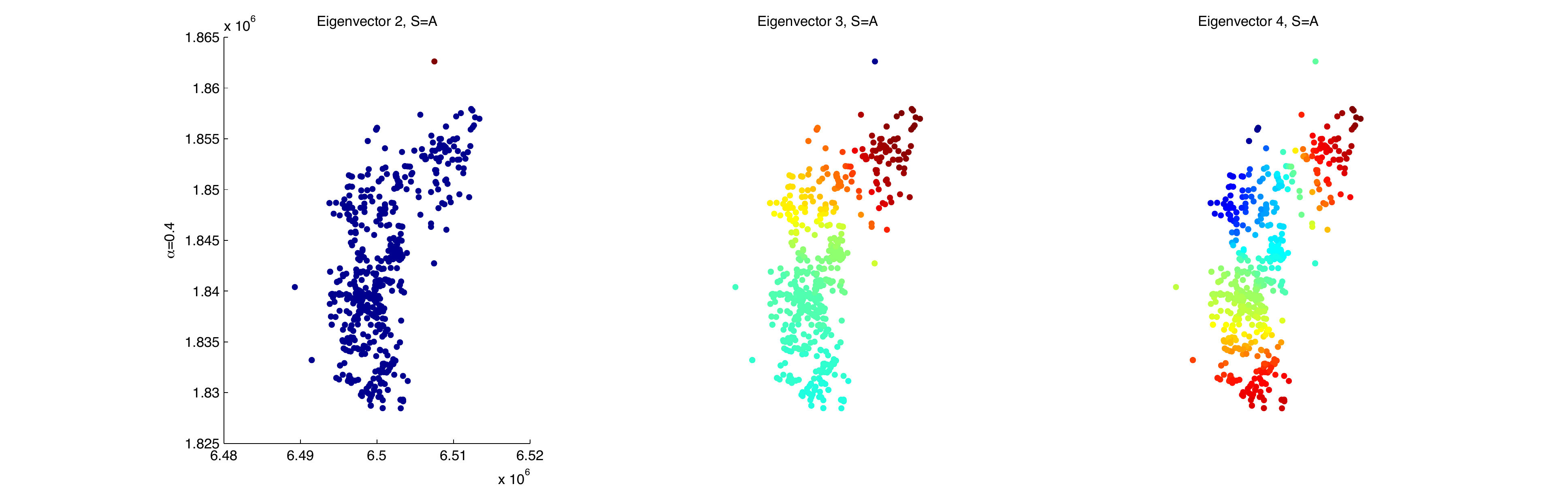}\\
		\includegraphics[width=\textwidth]{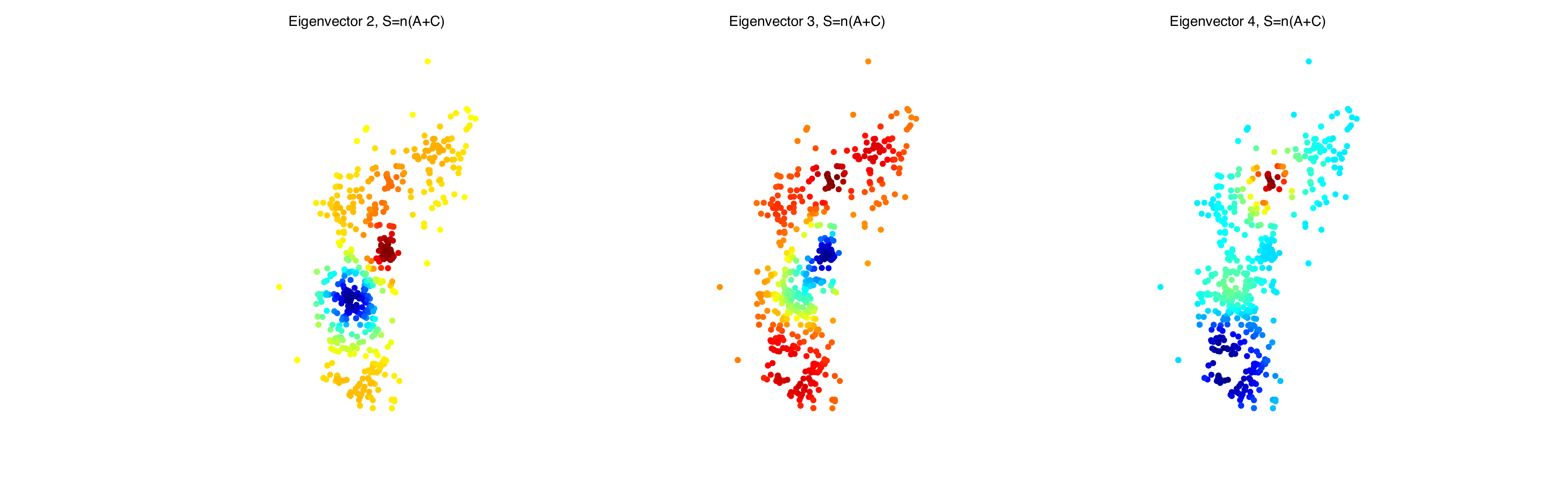}\\
		\end{array}$
	\end{center}
	\caption{\footnotesize  \textbf{Top: } The second, third, and fourth eigenvector of $D^{-1} W$, with $S=A$ and $\alpha=0.4$. The axes in the left picture have unit $10^6$ feet ($304.8$ km) with respect to the same coordinate origin as in Figure~\ref{fig:piechartclustering}. The color coding covers different ranges: Top left 0 (blue) to 1 (red), top middle -0.103 (blue) to 0.091 (red), top right -0.082 (blue) to 0.072 (red). \textbf{Bottom: } The second, third, and fourth eigenvector of $D^{-1} W$, with $S=n (A+C)$ and $\alpha=0.4$. The color coding covers different ranges: Top left -0.082 (blue) to 0.065 (red), top middle -0.091 (blue) to 0.048 (red), top right -0.066 (blue) to 0.115 (red).}\label{fig:eigenvectors_AvsA+1}
\end{figure}

Figure~\ref{fig:eigenvectors_AvsA+1} shows the second, third, and fourth eigenvectors of $D^{-1} W$ (because of the normalization the first eigenvector is constant, corresponding to eigenvalue 1) for $\alpha=0.4$, both when $S=A$ and when $S=n (A+C)$ is used. We see that hotspots have appeared after our rank-one update (and renormalization) of the social matrix $S$. Similar hotspots result for other $\alpha \in (0,1)$. An explanation for this behavior can be found in the behavior of eigenvectors under rank-one matrix updates, \cite{BunchNielsenSorensen78,GuEisenstat94}. Appendix~\ref{app:rank_one} gives more details. Similar hotspots (and changes in the metrics; see below) occur if other choices for $S$ are made that turn the zero entries into nonzero entries, \textit{e.g.} $S_{i,j} = e^{A_{i,j}}$, $S_{i,j}=e^{E_{i,j}}$ or $S_{i,j} = e^{-\theta_{i,j}}$, where $\theta$ is the spectral angle \cite{HarsanyiChang94,YuhasGoetzBoardman92}.

An analysis of the metrics when $S=n(A+C)$ shows that most metrics do not change significantly. The exceptions to this are two of the metrics described in Appendix~\ref{app:othermetrics}: The optimal value of the Hausdorff distance decreases to approximately 1350 meters, and the optimal value of the related minimal distance $M$ does not change much, but is now attained for a wide range of nonzero $\alpha$, not just for $\alpha=1$. Most importantly, the averages of the purity stay the same and while the averages of the $z$-Rand score decrease a bit, they do so within the error margins given by the standard deviations. Hence, the appearance of hotspots is not indicative of a global improvement in the clustering.

We tested whether the hotspots can be used to find the gangs located at these hotspots. For example, the hotspot seen in eigenvectors 2 (red) and 3 (blue) in the bottom row of Figure~\ref{fig:eigenvectors_AvsA+1} seems to correspond to Big Hazard in the left picture of Figure~\ref{fig:Hollenbeckdata}. We reran the spectral clustering algorithm, this time requesting only 2 clusters as output of the $k$-means algorithm and only using the second, third, or fourth eigenvector as input. The clusters that are created in this way correspond to ``hotspots versus the rest", but they do not necessarily correspond to ``one gang vs the rest". In the case of Big Hazard it does, but when only the second eigenvector is used the individuals in the big blue hotspot get clustered together. This hotspot does not correspond to a single gang.  We hypothesize that there is an interesting underlying sociological reason for this behavior: In the area of the blue hotspot a housing project, where several gangs claimed turf, was recently reconstructed displacing resident gang members.  Yet, even with these individuals being scattered across the city they remain tethered to their social space which remains in their established territories. \cite{HUD98,Olivo01}

We conclude that, from the available FI card data, it is not possible to cluster the individuals into communities that correspond to the different gangs with very high accuracy, for a variety of interesting reasons. First the social data is very sparse. The majority of  individuals are only involved in a couple of stops and most stops involve only a couple of people. Also, some gangs are only represented by a few individuals in the data sets: There are two gangs with only two members in the data set and two gangs with only three members. Second, the social reality of Hollenbeck is such that individuals and social contacts do not always adhere to gang boundaries, as the hotspot example above shows.

That the social data is both sparse and noisy (compared to the gang ground truth, which may be different from the social reality in Hollenbeck), we can see when we compare the connections in the FI card social adjacency matrix $A$ with the ground truth connections (the ground truth connects all members belonging to the same gang and has no connections between members of different gangs). We then see that\footnote{Not counting the diagonal which always contains ones.} only 2.66\% of all the ground truth connections (intra-gang connections) are present in $A$. On the other hand 11.32\% of the connections that are present in $A$ are false positives, \textit{i.e.} they are not present in the ground truth (inter-gang connections). Because missing data in $A$ (contacts that were not observed) show up as zeros in $A$, it is not surprising that of all the zeros in the ground truth 99.98\% are present in $A$ and only 5.56\% of the zeros in $A$ are false negatives.

Another indication of the sparsity is the fact that on average each individual in the data we used is connected to only 1.2754 $\pm$ 1.8946 other people\footnote{This number is of course always nonnegative, even though the standard deviation is larger than the mean.}. The maximum number of connections for an individual in the data is 23, but 315 of the 748 gang members (42\%) are not connected to any other individual.

Future studies can focus on the question whether the false positives and negatives in $A$ are noise or caused by social structures violating gang boundaries, possibly by comparing the impure clusters with inter-gang rivalry and friendship networks \cite{TitaRileyRidgewayGrammichAbrahamseGreenwood03,RadilFlintTita10,ShortMohlerBrantinghamTita12}. Another possibility is that the false positives and negatives betray a flaw in our assumption that individuals that are stopped together have a friendly relationship. Because of the non-criminal nature of the stops, this seems a justified assumption, but it is not unthinkable that some people that are stopped together have a neutral or even antagonistic relationship.

To rule out a third possibility for the lack of highly accurate clustering results, namely limitations of the spectral clustering method, we will now study how the method performs on quasi-artificial data constructed from the ground truth.

\section{Stability of metrics}\label{sec:stability}

To investigate the effect of having less sparse social data we compute purity using $S=GT(p,q)$. $GT(p,q)$ is a matrix containing a fraction $p$ of the ground truth connections, a further fraction $q$ of which are changed from true to false positive to simulate noise. In a sense, $p$ indicates how many connections are observed and $q$ determines how many of those are between members of different gangs. The matrix $GT(p,q)$ for $p,q\in [0,1]$ is constructed from the ground truth as follows. Let $GT(1, 0)$ be the gang ground truth matrix, \textit{i.e.} it has entry $(GT(1,0))_{i,j}=1$ if and only if $i$ and $j$ are members of the same gang (including $i=j$). Next construct the matrix $GT(p,0)$ by uniformly at random changing a fraction $1-p$ of all the strictly upper triangular ones in $GT(1,0)$ to zeros and symmetrizing the matrix. Finally, make $GT(p,q)$ by uniformly at random changing a fraction $q$ of the strictly upper triangular ones in $GT(p,0)$ to zeros and changing the same number (not fraction) of randomly selected strictly upper triangular zeros to ones, and in the end symmetrizing the matrix again. In other words, we start out with the ground truth matrix, keep a fraction $p$ of all connections, and then change a further fraction $q$ from true positives into false inter-gang connections.

\begin{figure}[ht]
	\begin{center}
		$\begin{array}{c@{\hspace{.1in}}}
		\includegraphics[width=\textwidth]{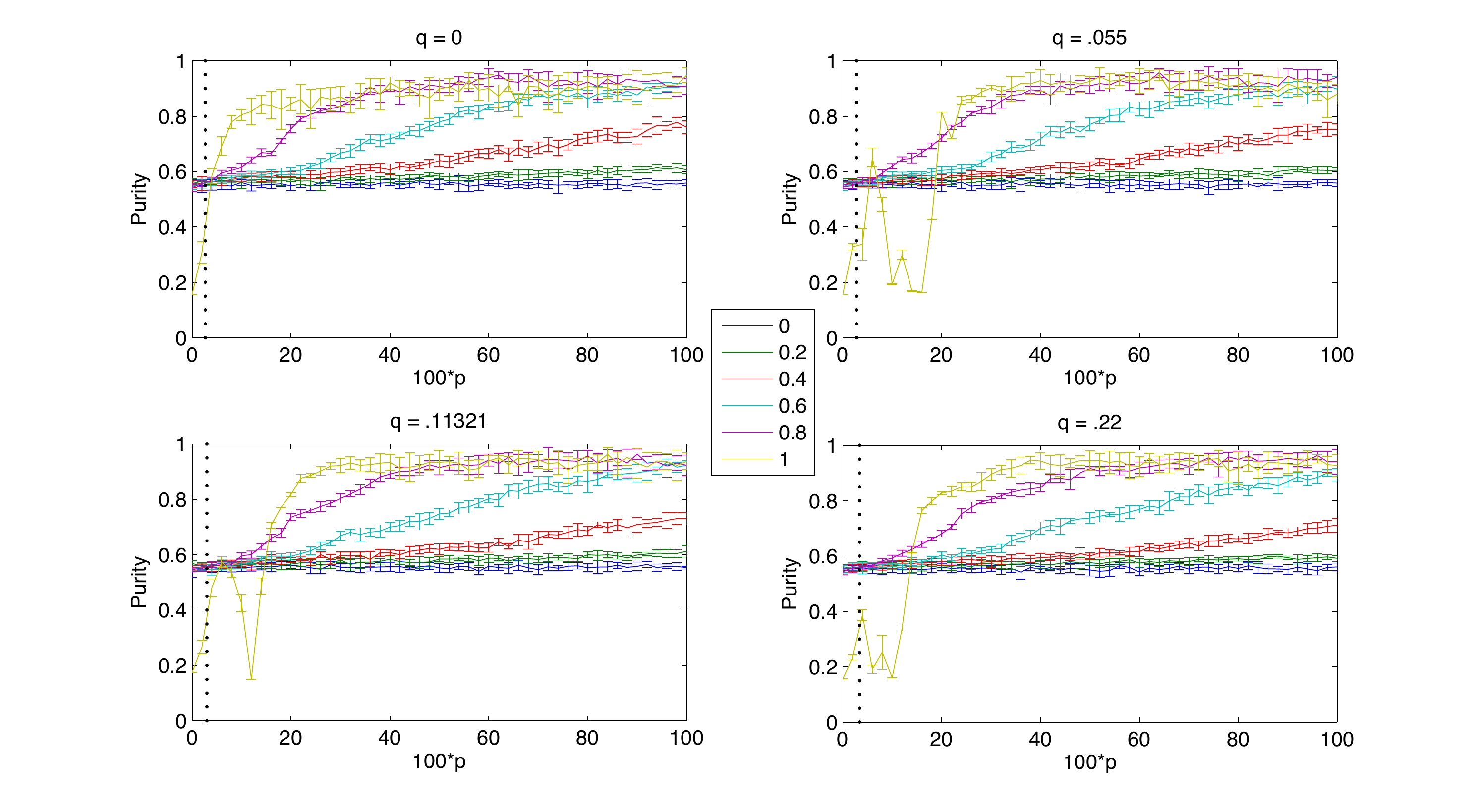}\\
		\end{array}$
	\end{center}
	\caption{\footnotesize Plots of the purity using $S=GT(p,q)$ for different values of $q$ (the different plots) and $\alpha$ (the different lines within each plot) for varying values of $p$. The plotted purity values per set of parameter values are averages over 10 $k$-means runs, the error bars are given by the standard deviation over these runs. The dotted vertical lines indicate the values of $p$ for which the number of true positives in $GT(p,q)$ is equal to the number of true positives in $A$.}\label{fig:varyingpq}
\end{figure}

In Figure~\ref{fig:varyingpq} we show the average purity over 10 $k$-means runs using $S=GT(p,q)$ for different values of $p$, $q$, and $\alpha$. To compare these results to the results we got using the observed social data $A$ from the FI card data set, we remember from Section~\ref{sec:socialmatrix} that $A$ contains only 2.66\% of the true intra-gang connections which are present in $GT(1,0)$. This roughly corresponds to $p$. On the other hand the total percentage of false positives (\textit{i.e.} inter-gang connections) in $A$ is $11.32\%$, roughly corresponding to $q$. By increasing $p$ and varying $q$ in our synthetic data $GT(p,q)$ we extend the observed social links, adding increased amounts of the true gang affiliations with various levels of noise (missing intra-gang social connections and falsely present inter-gang connections).

To investigate the effect of the police collecting more data at the same noise rate we keep $q$ fixed, allowing only the percentage of social links to vary. Low values of $\alpha$, \textit{e.g.} $\alpha=0$ and $\alpha=0.2$, show again that a baseline level of purity (about 56\%) is obtained by the geographical information only and hence is unaffected by changing $p$. As the noise level, $q$, is varied in the four plots in Figure~\ref{fig:varyingpq}, a general trend is clear: larger values of $0 \leq \alpha < 1$ correlate to higher purity values. This trend is enhanced as the percentage of social links in the network increases.  As expected, when only social information is used, $\alpha =1$, the algorithm is more sensitive to variations in the social structure.  This sensitivity is most pronounced at low levels, when the total percentage of social links are below 20.  Even at low levels of noise, $q=5.5$, using only social information is highly sensitive.  This suggests that $\alpha$ values strictly less than one are more robust to noisy links in the network.  The optimal choice of $\alpha=8$  here is more robust and consistently produces high purity values across the range of percentages of ground truth. A possible explanation for this sensitivity at $\alpha=1$ and the persistent dip in purity for this value of $\alpha$ and low values of $p$ is that for fixed $q$ and increasing $p$ the absolute (but not the relative) number of noisy entries increases. At low total number of connections these noisy entries wreak havoc on the purity in the absence of the mitigating geographical information. The bottom left of Figure~\ref{fig:varyingpq} shows a noise level of $q=0.11321$ which is set to match with what was obtained in the observed data.  The dotted vertical lines are plotted at values of $p$ satisfying
\[
p = \frac{\text{total number of true positives in $A$}}{\text{total number of upper triangular ones in } GT(1,0)} \frac1{1-q} = \frac{423}{15,904} \frac1{1-q}.
\]
For this value of $p$ the total number of true positives in $GT(p,q)$ is $\displaystyle 15,904 \cdot p \cdot (1-q)  = 423$ which is equal to the total number of true positives in $A$.

It is clear from the pictures that collecting and using more data (increasing $p$), even if it is noisy, has a much bigger impact on the purity than lowering the 11.32\% rate of false positives.

As remarked in Section~\ref{sec:socialmatrix} already we ran the same simulations using a social environment matrix like $E$ as choice for the social matrix $S$, but built from $GT(p,q)$ instead of $A$. The results were very similar to those using $S=GT(p,q)$ showing that also for less sparse data there does not appear to be much of a difference between using the social adjacency matrix or the social environment matrix. We also ran simulations computing the $z$-Rand score instead of purity using $S=GT(p,q)$. Again, the qualitative behavior was similar to the results discussed above.

\section{Conclusion and discussion}

In this paper we have applied the method of spectral clustering to an LAPD FI card data set concerning gang members in the policing area of Hollenbeck. Based on stop locations and social contacts only we clustered all the individuals into groups, that we interpret as corresponding to social communities. We showed that the geographical information leads to a baseline clustering which is about 56\% pure compared to the ground truth gang affiliations provided by the LAPD. Adding social data can improve the results a lot, if it is not too sparse. The data which is currently available is very sparse and improves only a little on the baseline purity, but our simulations show that improving the social data a little can lead to large improvements in the clustering. 

An extra complicating factor, which needs external data to be dealt with, is the very real possibility that the actual social communities in Hollenbeck are not strictly separated along gang lines. Extra sociological information, such as friendship or rivalry networks between gangs, can be used in conjunction with clustering method to investigate the question how much of the social structures observed in Hollenbeck are the results of gang membership.

Future studies will also investigate the effect of using different methods, including the multislice method of \cite{MuchaRichardsonMaconPorterOnnela10}, the alternative spectral clustering method of \cite{GhoshLermanSurachawalaVoevodskiTeng11,GhoshLerman12} based on an underlying non-conservative dynamic process (as opposed to a conservative random walk), and the nonlinear Ginzburg-Landau method of \cite{BertozziFlenner12}, which uses a few known gang affiliations as training data. The question how partially labeled data helps with clustering in a semi-supervised approach was explored in \cite{AllahverdyanVerSteegGalstyan10}.

\bigskip

\textbf{Acknowledgements.} The FI card data set used in this work was collected by the LAPD Hollenbeck Division and digitized, scrubbed, and anonymized for use by Megan Halvorson, Shannon Reid, Matthew Valasik, James Wo, and George E. Tita, at the Department of Criminology, Law and Society of UCI. The data analysis work was started by (then) (under)graduate students Raymond Ahn, Peter Elliott, and Kyle Luh, as part of a project in the 2011 summer REU program in applied mathematics project at UCLA organized by Andrea L. Bertozzi. The project's mentors, Yves van Gennip and Blake Hunter, together with P. Jeffrey Brantingham, extended the summer project into the current paper. We thank Matthew Valasik, Raymond Ahn, Peter Elliott, and Kyle Luh, for their continued assistance on parts of the paper, and Mason A. Porter of the Oxford Centre for Industrial and Applied Mathematics of the University of Oxford for a number of insightful discussions. This work was made possible by funding from NSF grant DMS-1045536, NSF grant DMS-0968309, ONR grant N000141010221, AFOSR MURI grant FA9550-10-1-0569 and ONR grant N000141210040.

\appendix

\section{Other metrics}\label{app:othermetrics}

In some cases it is useful to look beyond purity and the $z$-Rand score which we discussed in Sections~\ref{sec:metrics} and~\ref{sec:performance}. Hence we also define metrics that measure the gang homogeneity within clusters, the gang heterogeneity between clusters, and the accuracy of the geographical placement of our clusters. To give an impression of how our data performs for these metrics, we give the order of magnitude of their typical values observed as averages over 10 $k$-means runs.

Recall from Section~\ref{sec:metrics} that $w_{11}$ is the number of pairs which belong both to the same cluster in our $k$-means clustering and to the same gang. Analogously $w_{10}$, $w_{01}$, and $w_{00}$ are the numbers of pairs which are in the same $k$-means cluster but different gangs, different $k$-means clusters but the same gang, and different $k$-means clusters and different gangs respectively, \textit{e.g.} \cite{Meila07,TraudKelsicMuchaPorter11} and references therein. 

Considering the error bars, the choice of $\alpha$ does not matter too much for $w_{11} \approx 6,000$ and $w_{01} \approx 9,800$. As long as $\alpha<1$ it also does not matter much for $w_{10} \approx 10,000$ and $w_{00} \approx 250,000$.

We define ingroup homogeneity as the probability of choosing two individuals belonging to the same gang if we first randomly pick a cluster (with equal probability) and then randomly choose two people from that cluster. We also define a scaled ingroup homogeneity, by taking the probability of choosing a cluster proportional to the cluster size. Analogously we define the outgroup heterogeneity as the probability of choosing two individuals belonging to different gangs if we first pick two different clusters at random and then choose one individual from each cluster. The scaled outgroup heterogeneity again weights the probability of picking a cluster by its size. 

We see a sharp drop in ingroup homogeneity when going from the unscaled ($\approx 0.58$) to the scaled ($\approx 0.40$) version, indicating the presence of a lot of small clusters, which are likely to be very homogeneous, but have a small chance of being picked out in the scaled version. This effect is not present for the outgroup heterogeneity ($\approx 0.96$ for either the scaled or unscaled version) because the small cluster effect is tiny compared to the overall heterogeneity.

We also compare the centroids of our clusters (the average of the positions of all individuals in a cluster) in space to the centroids based on the true gang affiliations. The Hausdorff distance is the maximum distance one has to travel to get from a cluster centroid to its nearest gang centroid or vice versa. We define $M$ as the average of these distances, instead of the maximum. For comparison, the maximum distance between two individuals in the data set is 10,637 meters. 

The Hausdorff distance ($\approx 2200$ meters) does not change much with $\alpha$ (but the standard deviation is very large when $\alpha=1$). Surprisingly the average distance $M$ is minimal ($\approx 450$ meters) for $\alpha=1$, about 100 meters less compared to $\alpha<1$. The large difference between $M$ and the Hausdorff distance for any $\alpha$ indicates most centroids are clustered close together, but there are some outliers.

The cluster distance (code from \cite{CoenAnsariFillmore10}) computes the ratio of the optimal transport distance between the centroids of our clustering and the ground truth and a naive transport distance which disallows the splitting up of mass. The underlying distance between centroids is given by the optimal transport distance between clusters. This distance ranges between 0 and 1, with low values indicating a significant overlap between the centroids.
The cluster distance ($\approx 0.29$) is significantly better if $\alpha<1$, showing a significant geographic overlap between the spectral clustering and the clustering by gang.

\section{Different number of clusters}\label{app:different}

\begin{figure}[ht]
	\begin{center}
		$\begin{array}{c@{\hspace{.1in}}}
		\includegraphics[width=0.7\textwidth]{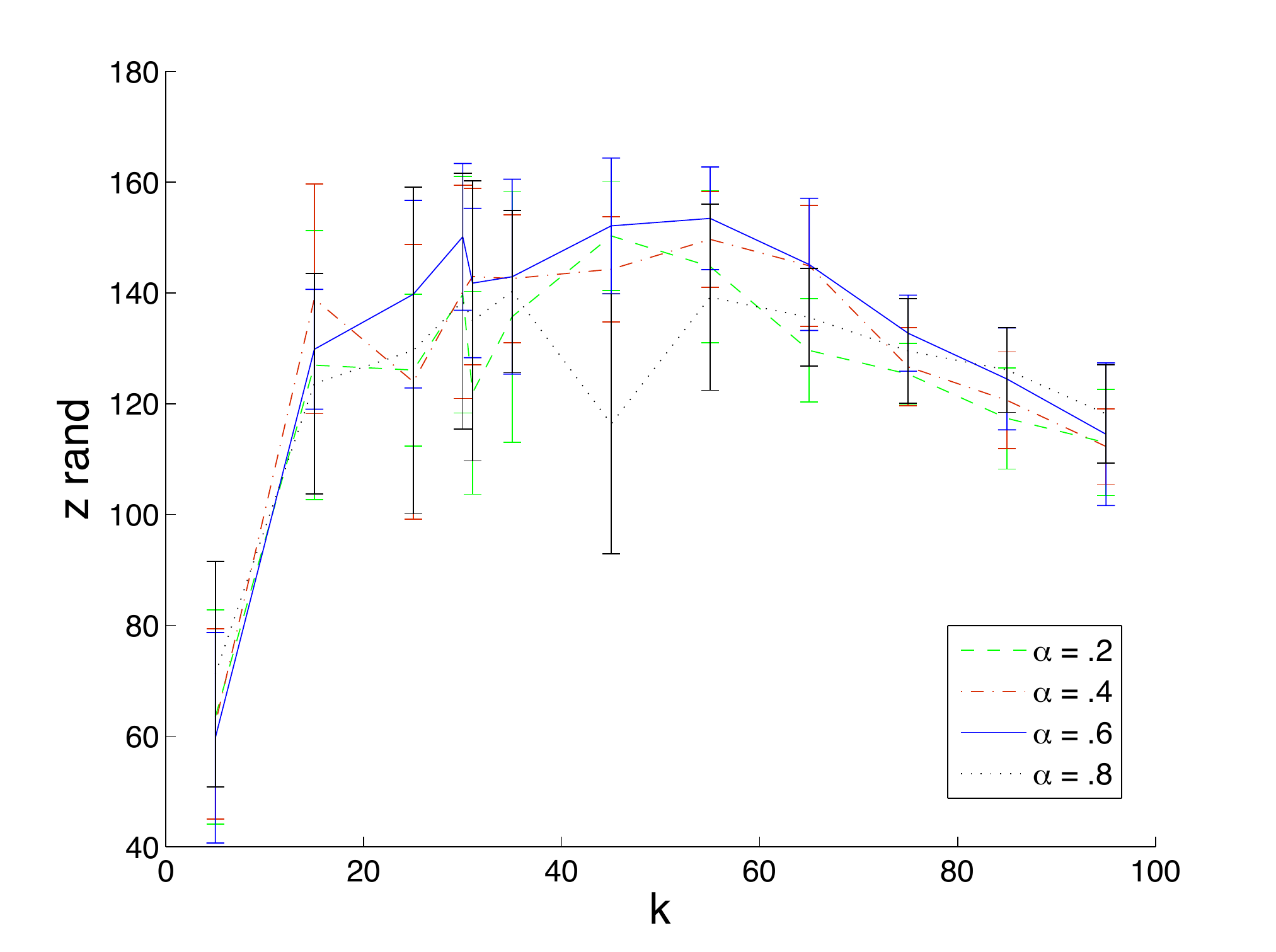}\\
		\end{array}$
	\end{center}
	\caption{\footnotesize  The mean $z$-Rand score over 10 $k$-means runs, plotted against different values of $k$. The different lines correspond to different values for $\alpha\in \{0.2, 0.4, 0.6, 0.8\}$. The error bars indicate the standard deviation.}\label{fig:differentks}
\end{figure}

In this section we briefly discuss results obtained for values of $k$ different from 31. Note that most of the metrics discussed in Section~\ref{sec:metrics} and Appendix~\ref{app:othermetrics} are biased towards having either more or fewer clusters. For example, as discussed in Section~\ref{sec:metrics}, purity is biased towards more clusters. Indeed, we computed the values of all the metrics for $k\in \{5, 25, 30, 35, 60\}$ and noticed that the biased metrics behave as {\it a priori} expected, based on their biases. This means most of the metrics are bad choices for comparing results obtained for different values of $k$. The exception to this is the $z$-Rand score, which does allow us to compare clusterings at different values of $k$ to the gang affiliation ground truth. We computed the $z$-Rand scores for clusterings obtained for a range of different values of $k$, between 5 and 95. The results can be seen in Figure~\ref{fig:differentks}.

As can be seen from this figure, the $z$-Rand has a maximum around $k=55$, although most $k$ values between about 25 and 65 give similar results, within the range of one standard deviation. We see that, as measured by the $z$-Rand score, the quality of the clustering is quite stable with respect to $k$.

\section{Rank-one matrix updates}\label{app:rank_one}

Here we give details explaining how the eigenvectors of a symmetric matrix $W$ change when we add a constant matrix. Assume for simplicity\footnoteremember{more complicated}{Note that what we are doing in our simulations is slightly more complicated: We use $\alpha n (S+C) + (1-\alpha) e^{-d_{i,j}^2/\sigma^2}$, so in addition to adding a constant matrix $S$ is multiplied by a normalization factor $n=(\underset{i,j}\max\, (S_{i,j} + 1))^{-1}$.} that we want to know the eigenvalues of $W+C$, where $C$ is an $N$ by $N$ ($N=748$) matrix whose entries $C_{i,j}$ are all 1. Let $Q$ be a matrix that has as $i^{\text{th}}$ column the eigenvector $v_i$ of $W$ with corresponding eigenvalue $d_i$. Let $D$ be the diagonal matrix containing these eigenvalues, then we have the decomposition $W = QDQ^T$. Write $b$ for the $N$ by 1 vector with entries $b_i = 1$, such that $C = b b^T$. If we write $z:= Q^{-1} b$ then
\[
W+C = Q (D + z z^T) Q^T = Q (X \Lambda X^T) Q^T,
\]
where $X$ has the $i^\text{th}$ eigenvector of $D+zz^T$ as $i^{\text{th}}$ column and $\Lambda$ is the diagonal matrix with the corresponding eigenvalues $\lambda_i$. We are interested in $QX$, which is the matrix containing the eigenvectors of $W+C$. According to \cite{BunchNielsenSorensen78} and \cite[Lemma 2.1]{GuEisenstat94}\footnote{In order to use this result we need to assume that all the eigenvalues $d_i$ are simple, \textit{i.e.} $W$ should have different eigenvalues. This might not be a completely true assumption in our case, although it typically holds for most eigenvalues unless $W$ has a well separated block diagonal structure.} we have for the $i^{\text{th}}$ column of $X$:
\[
X_{:,i} = c_i \left(\frac{z_1}{d_1-\lambda_i}, \ldots, \frac{z_N}{d_N-\lambda_i}\right)^T,
\]
with normalization constant $c_i = \sqrt{\sum_{j=1}^N \frac{z_j^2}{(d_j-\lambda_i)^2}}$.

Now
\begin{align*}
(QX)_{k,i} &= Q_{k,:} \cdot X_{:,i} = Q_{k,:} \cdot c_i \left( Q^{-1}_{1,:} \cdot b/(d_1-\lambda_i), \ldots, Q^{-1}_{N,:} \cdot b/(d_N-\lambda_i) \right)^T\\
&= c_i \sum_{l,m=1}^N \frac{Q_{k,l} Q^{-1}_{l,m} b_m}{d_l-\lambda_i}.
\end{align*}
Since $b_m = 1$ for all $m$ we have $(QX)_{k,i} = c_i \sum_{m=1}^N (Q F Q^{-1})_{k,m}$ where $F$ is the diagonal matrix with entries $F_{ll} = \frac1{d_l-\lambda_i}$. Since $Q$ has the eigenvectors $v_l$ as columns and $Q^{-1}$ is its transpose we conclude
\[
(QX)_{k,i} = c_i\sum_{m=1}^N \left[ (v_1, \ldots, v_N) \left(\frac1{d_1-\lambda_i} v_1, \ldots, \frac1{d_N-\lambda_i} v_N\right)^T \right]_{k,m} = c_i \sum_{m,l=1}^N (v_l)_k \frac1{d_l-\lambda_i} (v_l)_m.
\]
Finally, since the eigenvectors $v_l$ are normalized we find that the $k^{\text{th}}$ component of the $i^{\text{th}}$ new eigenvector is given by
\[
(QX)_{k,i} = c_i \sum_{l=1}^N \frac{(v_l)_k}{d_l-\lambda_i}.
\]

Also, according to \cite[Theorem 1]{BunchNielsenSorensen78}, the eigenvalues $\lambda_i$ are given by
\[
\lambda_i = d_i + N^2 \mu_i,
\]
for some $\mu_i \in [0,1]$ which satisfy $\sum_{i=1}^N \mu_i=1$.

\begin{figure}[ht]
	\begin{center}
		$\begin{array}{c@{\hspace{.1in}}}
		\includegraphics[width=0.7\textwidth]{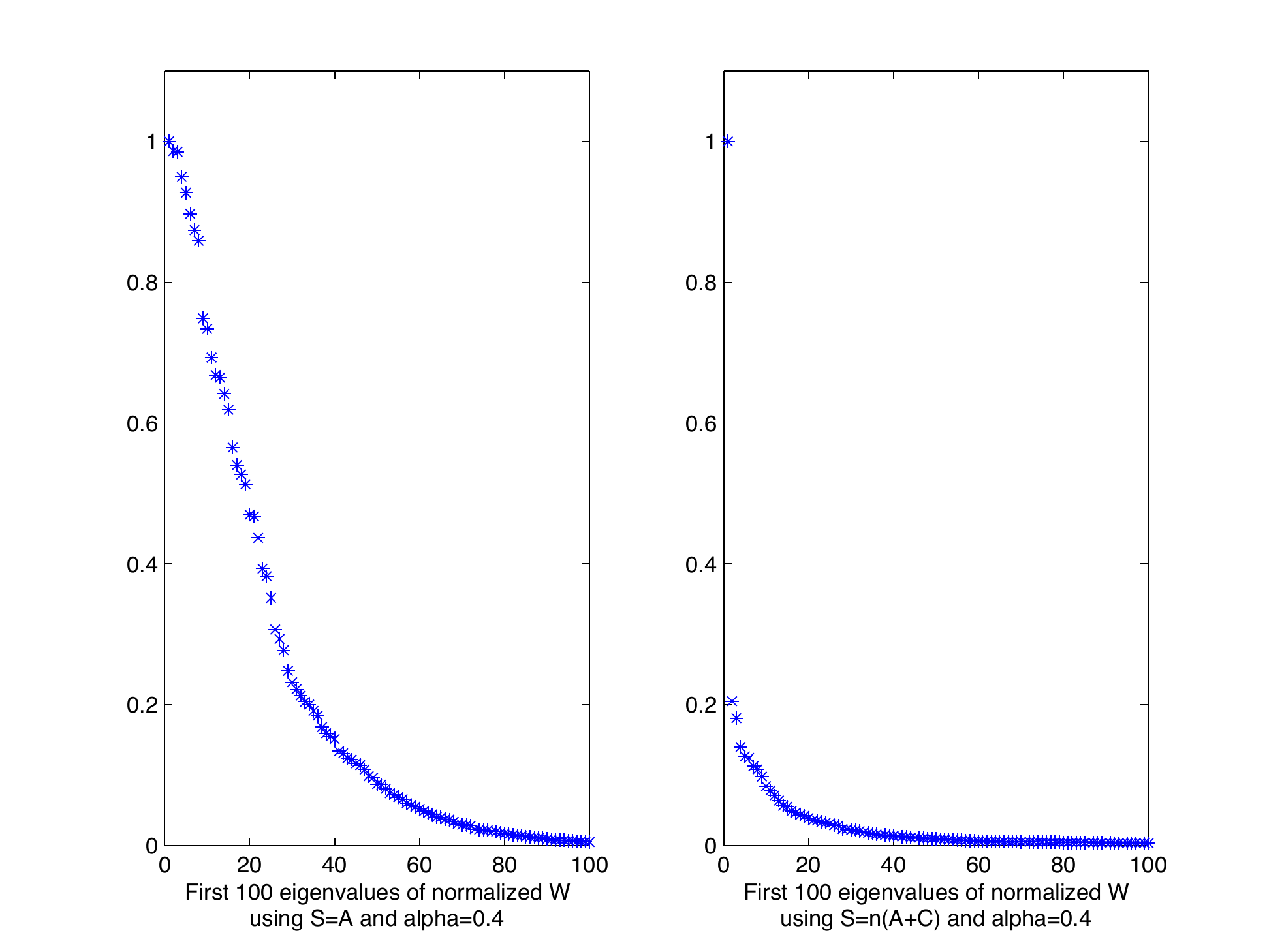}\\
		\end{array}$
	\end{center}
	\caption{\footnotesize  \textbf{Left: } The first 100 eigenvalues of $D^{-1} W$, with $S=A$ and $\alpha=0.4$. \textbf{Bottom: } The first 100 eigenvalues of $D^{-1} W$, with $S=n (A+C)$ and $\alpha=0.4$. }\label{fig:eigvals}
\end{figure}

If we apply this idea to our geosocial eigenvectors, we see in Figure~\ref{fig:eigvals} that most of the eigenvalues of $W$ and $W+C$\footnoterecall{more complicated} are close to zero and hence close to each other. Only among the first couple dozen there are large differences. This means that most of the new eigenvectors are more or less equally weighted sums of all the old eigenvectors belonging to the small eigenvalues and hence lose most structure. It is therefore up to the relatively few remaining eigenvectors (those corresponding to the larger eigenvalues) to pick up all the relevant structure. This might be an explanation of why hotspots appear.

\bibliographystyle{acm}
\bibliography{bibliography}

\end{document}